\documentclass[preprint2]{aastex}

\slugcomment{Revised 24 June 2002; Accepted July 8}

\shorttitle{Spectroscopic Detection of a Protostellar Photosphere}
\shortauthors{Greene and Lada}

\begin{document}

\title{Spectroscopic Detection of a Stellar-like Photosphere
in an Accreting Protostar\altaffilmark{1}}


\author{Thomas P. Greene}
\affil{NASA's Ames Research Center, M.S. 245-6, Moffett Field, CA 
94035-1000}
\email{tgreene@arc.nasa.gov}

\author{Charles J. Lada}
\affil{Harvard-Smithsonian Center for Astrophysics\\ 60 Garden Street,
Cambridge, MA 02138}
\email{clada@cfa.harvard.edu}

\altaffiltext{1}{Data presented herein were obtained at the W.M. Keck
Observatory, which is operated as a scientific partnership among the
California Institute of Technology, the University of California and
the National Aeronautics and Space Administration. The Observatory
was made possible by the generous financial support of the W.M. Keck
Foundation.}

\begin{abstract}
 We present high-resolution ($R \simeq 18,000$), 
high signal-to-noise
2 $\mu$m spectra of two luminous, X-ray flaring Class I protostars in
the $\rho$ Ophiuchi cloud acquired with the NIRSPEC
spectrograph of the Keck II telescope. We present the first spectrum 
of a highly veiled, strongly
accreting protostar which shows photospheric absorption
features and demonstrates the stellar nature of its central
core. We find the spectrum of the luminous ($L_{bol}$ = 10 $L_\odot$)
protostellar source, YLW 15, to be stellar-like with numerous atomic 
and molecular absorption features, indicative of 
a K5 IV/V spectral type and a continuum veiling $r_k = 3.0$. Its derived
stellar luminosity (3~$L_\odot$) and stellar radius (3.1~$R_\odot$) are
consistent with those of a 0.5 $M_\odot$ pre-main-sequence star.
However, 70\% of its bolometric luminosity is due to mass accretion,
whose rate we estimate to be $1.6 \times 10^{-6} M_\odot$ yr$^{-1}$
onto the protostellar core.
We determine that excess infrared emission produced by the circumstellar
accretion disk, the inner infalling envelope, and accretion shocks 
at the surface of the stellar core of YLW 15 all
contribute significantly to its near-IR continuum veiling. 
Its projected rotation velocity $v$ sin $i$ = 50 km s$^{-1}$ is
comparable to those of flat-spectrum protostars but considerably higher
than those of classical T Tauri stars in the $\rho$ Oph cloud.
The protostar may be magnetically coupled to its circumstellar disk
at a radius of 2 $R_*$. It is also plausible that this
protostar can shed over half its angular momentum and 
evolve into a more slowly rotating classical T Tauri 
star by remaining coupled to its circumstellar disk (at increasing 
radius) as its accretion rate drops by an order of magnitude during the
rapid transition between the Class I and Class II phases of evolution.

The spectrum of WL 6 does not show any photospheric
absorption features, and we estimate that its continuum veiling is $r_k
\geq 4.6$. Its low bolometric luminosity (2 $L_{\odot}$) and high veiling 
dictate that its central protostar is very low mass, M $\sim$ 0.1 M$_\odot$.

\end{abstract}



\section{Introduction}

The discovery of accreting protostars has been a triumph for understanding
low-mass star formation. Infrared and
sub-millimeter data have revealed that these objects are deeply
embedded, cold, and luminous. The spectral energy distributions (SEDs)
of such Class I protostars gradually rise from the near-IR, 
peak in the far-IR ($\lambda \sim 100 \mu$m), and fall off in the
sub-mm and mm wavelength regimes \citep[e.g.,][]{ALS87}. Observations
and models have indicated that protostars consist of three primary
components: a massive infalling envelope, a circumstellar accretion
disk, and an embryonic stellar core. 

With a size of $\sim$ 10$^4$ AU,
the infalling envelope is the largest component of a protostellar 
object. It extends to within a few AU of the embryonic 
stellar core and absorbs and reprocesses all the visible (and UV) as
well as much
of the near-infrared radiation emitted by the central protostar. 
The infalling envelope contributes 
the bulk of the observed far-infrared to millimeter-wave luminosity and a
significant portion of the detected near- and mid-infrared emission. 

The circumstellar accretion disk can be as much as a
few hundred AU in size and it
contributes a significant portion of the observed mid- and near-infrared
luminosity of the protostar. During the protostellar phase of 
evolution, the embryonic stellar core at its center
grows primarily through accretion of 
material from the circumstellar disk, which itself acquires its mass
directly from the infalling envelope.

The embryonic stellar core
at the center of the protostar is thought to be a stellar-like object
a few solar radii in extent \citep[e.g.,][]{SST80}. It is the source
of the bulk of the protostellar luminosity which is generated by both
a strong accretion shock at the stellar surface and the radiation of
internal stellar energy. This internal energy is provided by 
its self-gravity as well as the release of gravitational 
potential energy during the protostar's assembly from accreting and infalling
material. The core's accretion-dominated luminosity is radiated primarily
at UV and visual
wavelengths, but most, if not all of this radiation is absorbed 
by dust in the infalling envelope which renders the central protostar
invisible at these wavelengths. As a result, 
little is known about the nature of the embryonic stellar cores at
the hearts of protostellar objects.  

The observational technique of high-resolution near-IR spectroscopy has the
potential to reveal the nature of these heavily embedded central
objects. Indeed, we have recently been able to demonstrate the stellar nature
of the central objects in flat-spectrum protostars and thus 
discern their spectral types, effective
temperatures, rotation velocities, luminosity classes, and masses
using this technique \citep[hereafter Paper I]
{PaperI}.  These objects are in a later evolutionary state than Class I
protostars; they are still surrounded by substantial circumstellar disks and
envelopes but their accretion luminosities (and rates) have decreased below
Class I levels.  However, our previous high-resolution IR spectroscopic study
of Class I young stellar objects (YSOs)  did not reveal any
spectral absorption features in these protostars \citep[hereafter Paper
II]{PaperII}.  \cite{LR99} did detect absorption features in the
moderate-resolution near-IR spectrum of at least one Class I YSO (YLW 16A /
IRS 44), but the moderate continuum veiling of that object ($r_k = 1$)
suggests that it is intrinsically similar to a flat-spectrum YSO.
Also, \citet{KBTB98} detected TiO
absorptions in moderate-resolution optical spectra of three Class I YSOs in
the Tau-Aur dark clouds.  However, these are all low luminosity objects, $L_*$
$\simeq$ $L_{bol} \lesssim 0.5$ $L_\odot$, indicating relatively low
levels of accretion.  Thus there is very little or no
existing information concerning the nature and basic physical 
characteristics of the central stellar cores of protostars undergoing
significant mass accretion.

Observational determination of the effective temperatures and radii
(or surface gravities) of these deeply embedded central objects
would provide critical constraints for theoretical protostellar models 
which predict the values of these parameters as a
protostar evolves \citep{SST80}. Empirical knowledge concerning
the initial angular momentum of these objects as well as
the evolution of angular momentum in the earliest phases of young
stellar evolution is also highly desirable to test protostellar 
theory. Many classical T Tauri stars (CTTSs) embedded in the
$\rho$ Oph cloud and elsewhere are known to rotate slowly, $v$~sin
$i$~$<$~20~km~s$^{-1}$. However, we have found that flat-spectrum
protostars in this cloud have similar radii but rotate much more quickly, 
$v$~sin~$i$~$\simeq$~40 km s$^{-1}$ (Paper I). Do heavily-accreting
Class I protostars in Ophiuchus rotate as fast or faster? If so, how
is stellar angular momentum reduced so quickly by the classical T Tauri
pre-main-sequence (PMS) evolutionary phase? 

We have undertaken a new, very sensitive near-IR spectroscopic study
of Class I protostars with the 10 m Keck II telescope
in order to address these issues. The Keck observations are considerably
more sensitive than our previous deep infrared spectral survey made
with the 3-m NASA IRTF. In this paper
we report our findings for two of these objects in the $\rho$ Oph dark
cloud, YLW 15 (IRS 43) and WL 6. These are 2 of the 3 $\rho$ Oph Class I
protostars detected in hard X-rays by the ASCA satellite, and both
showed strong, variable, and flaring emission \citep{KKTY97}. We
describe our data acquisition and reduction in \S 2, analyze the spectra
in \S 3, discuss the results in \S 4, and summarize our conclusions in
\S 5. 

\section{Observations and Data Reduction}

Near-IR spectra of the $\rho$ Oph protostar YLW 15 were acquired
on 2001 July 7 UT, and the protostar WL 6 was observed on 2000 May 30
UT. MK spectral standards were also observed 2000 May 29 -- 30 UT.
All data were acquired with the 10-m Keck II telescope on Mauna Kea, 
Hawaii, using the NIRSPEC facility multi-order cryogenic echelle spectrograph 
\citep{McLeanetal98}.  Spectra were acquired with a 0\farcs58 (4 
pixel) wide slit, providing spectroscopic resolution 
$R \equiv \lambda / \delta \lambda$ = 18,000 (16.7 km s$^{-1}$).   
The plate scale was 0\farcs20 pixel$^{-1}$ along the 12$\arcsec$ slit 
length, and the seeing was 0\farcs5 -- 0\farcs6. The NIRSPEC gratings
were oriented to allow both the
2.21 $\mu$m Na and the 2.30 $\mu$m CO band head regions to fall onto the
instrument's 1024 $\times$ 1024 pixel InSb detector array, and its 
NIRSPEC-7 blocking filter was used. NIRSPEC was configured to acquire
simultaneously multiple cross-dispersed echelle orders 31 -- 36
(2.08 -- 2.45 $\mu$m, non-continuous) in 2000 May and orders 32 -- 38
(1.97 -- 2.38 $\mu$m, non-continuous) on 2001 July 7. Each order had
an observed spectral range $\Delta \lambda \simeq \lambda / 67$
($\Delta v \simeq$ 4450 km s$^{-1}$).

The internal instrument slit rotator was used to maintain a fixed
position angle on the sky when observing WL 6 and the MK standards on 2000 
May 29 -- 30. The slit was held physically stationary and thus
allowed to rotate on the sky (as the non-equatorially-mounted telescope
tracked) when observing on 2001 July 7 (YLW 15). 
Data were acquired in pairs of exposures of durations from less than
1 second (for giant MK standards) to up to 600 s (for the YSOs) each, 
with the telescope nodded $6\arcsec$ along the slit between frames 
so that object spectra were acquired in all exposures. WL 6 was
observed for a total of 30.0 minutes, while YLW 15 was observed for
73.3 minutes. Early-type (B7 -- A2) dwarfs were observed for
telluric correction of the MK standard
stellar spectra. HR 6054 (B7 IV) and HR 6070 (A0 V) were observed for
telluric correction of the WL 6 and YLW 15 spectra, respectively.
The telescope was automatically guided with frequent images from the
NIRSPEC internal SCAM IR camera during all exposures of more than several
seconds duration.  Spectra of the internal NIRSPEC continuum lamp
were taken for flat fields, and exposures of the its Ar, Ne, Kr, 
and Xe lamps were used for wavelength calibrations.

All data were reduced with IRAF. First, object and sky frames were 
differenced and then divided by flat fields.  Next, bad pixels were 
fixed via interpolation, and spectra were extracted with the APALL 
task.  Extracted spectra were typically 4 pixels (0\farcs6) wide 
along the slit (spatial) direction at their half-intensity points.  Spectra 
were wavelength calibrated using low-order fits to lines in the arc 
lamp exposures, and spectra at each slit position of each object were 
co-added.  Instrumental and atmospheric features were removed by 
dividing wavelength-calibrated object spectra by spectra of early-type 
stars observed at similar airmass at each slit position.  Combined 
spectra were produced by summing the spectra of both slit positions 
for each object and then shifting the spectra in wavelength to correct
for radial velocities. The standard spectra were shifted so that their
lines matched laboratory wavelengths, while the YSO spectra were
corrected for all Earth and solar system radial motions relative to the
local standard of rest. The YSO spectra were not further corrected for
the radial velocity of the $\rho$ Oph cloud. Finally, YSO spectra were
multiplied by an artificial spectrum of a T=10,000K black body in
order to restore their true continuum slopes (although this was a
small correction of under 10\% per order).

\section{Data Analysis and Results}

\subsection{Spectra}

The spectra of several MK standard stars are shown in Figure 1. The three
NIRSPEC orders with the most numerous absorption features are
displayed. These features are similar to ones seen by \cite{WH96} 
in stars of similar spectral type, although we present a finer grid
of spectra in the K2V -- M2V range. Spectra of these stars are
dominated by strong features of Si (2.0923 $\mu$m), Mg (2.1066 $\mu$m),
and Al (2.1099 $\mu$m) in the shortest wavelength order, with ratios 
Al/Mg and Al/Si being very sensitive to spectral type. Some of the strongest
lines in the next order are Na (2.2062, 2.2090 $\mu$m), Sc (2.2058,
2.2071 $\mu$m), and Ti (primarily 2.2217 and 2.2240 $\mu$m). The Ti lines
are very good indicators of spectral type in this range,
growing stronger from earlier to later types. The 2.2069 $\mu$m Si
line is also a good indicator of spectral type, weakening from early to
later types. Therefore the ratio of Ti/Si is a particularly good
spectral type diagnostic in this order. The strongest features in the
longest wavelength order are Mg (2.2814 $\mu$m) and the CO $v = 0 -
2$ band head and rotation-vibration lines (2.2935 $\mu$m and
long-wards). The Mg line becomes significantly weaker in later
spectral types. All transitions identified above and hereafter 
(and in Fig. 1) occur in neutral species.

\begin{figure}
\figurenum{1}
\plotone{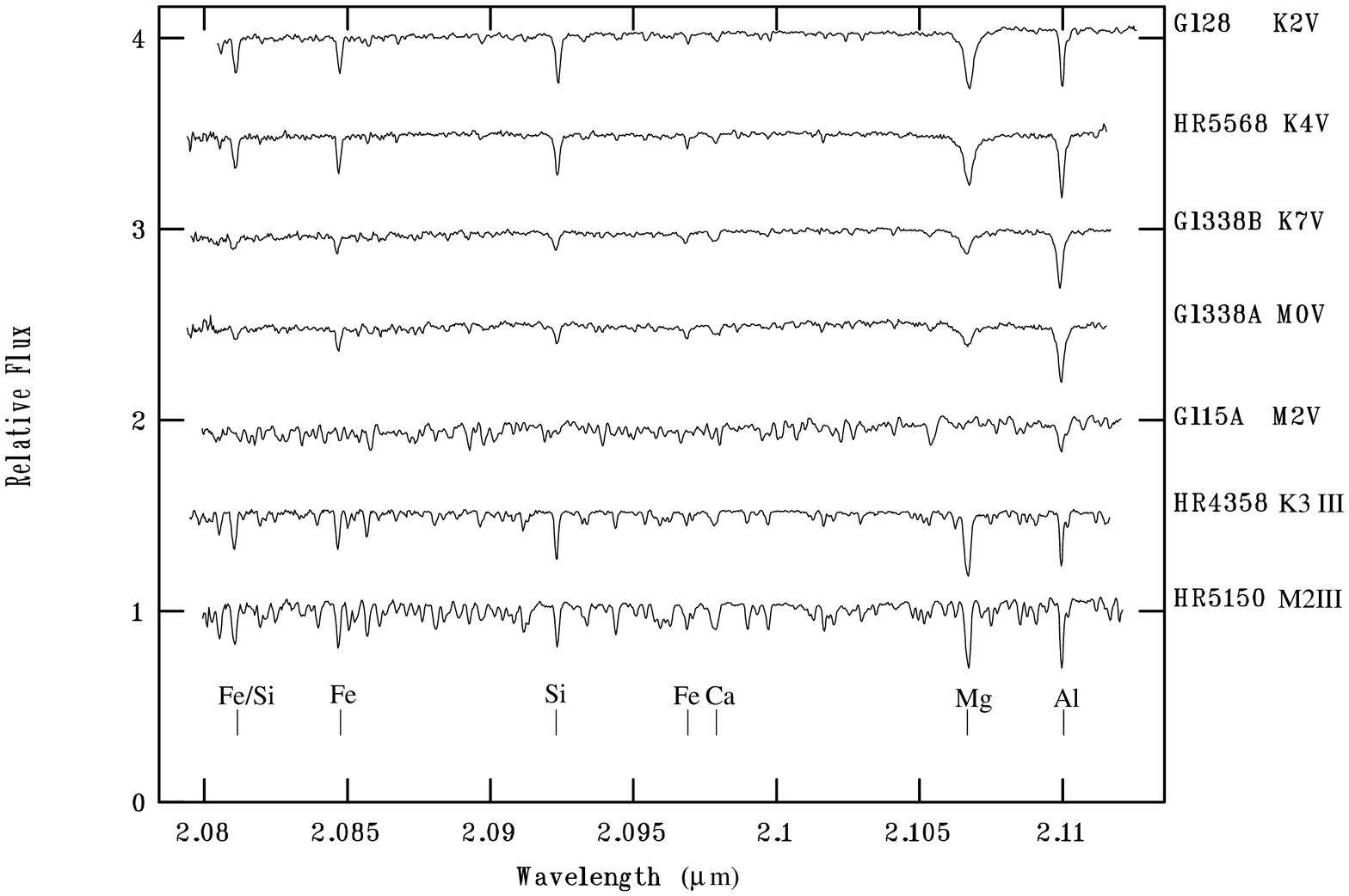}
\plotone{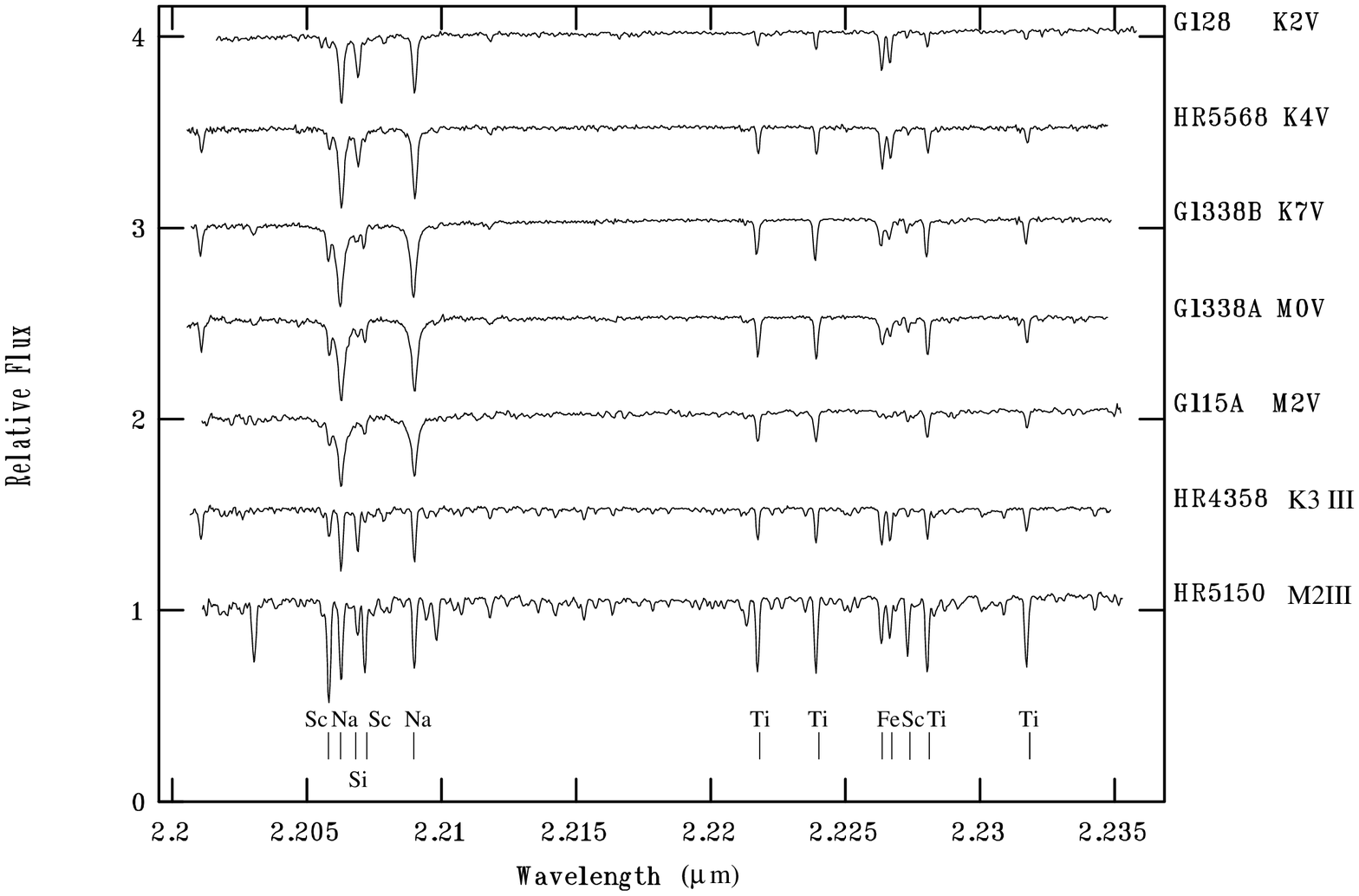}
\plotone{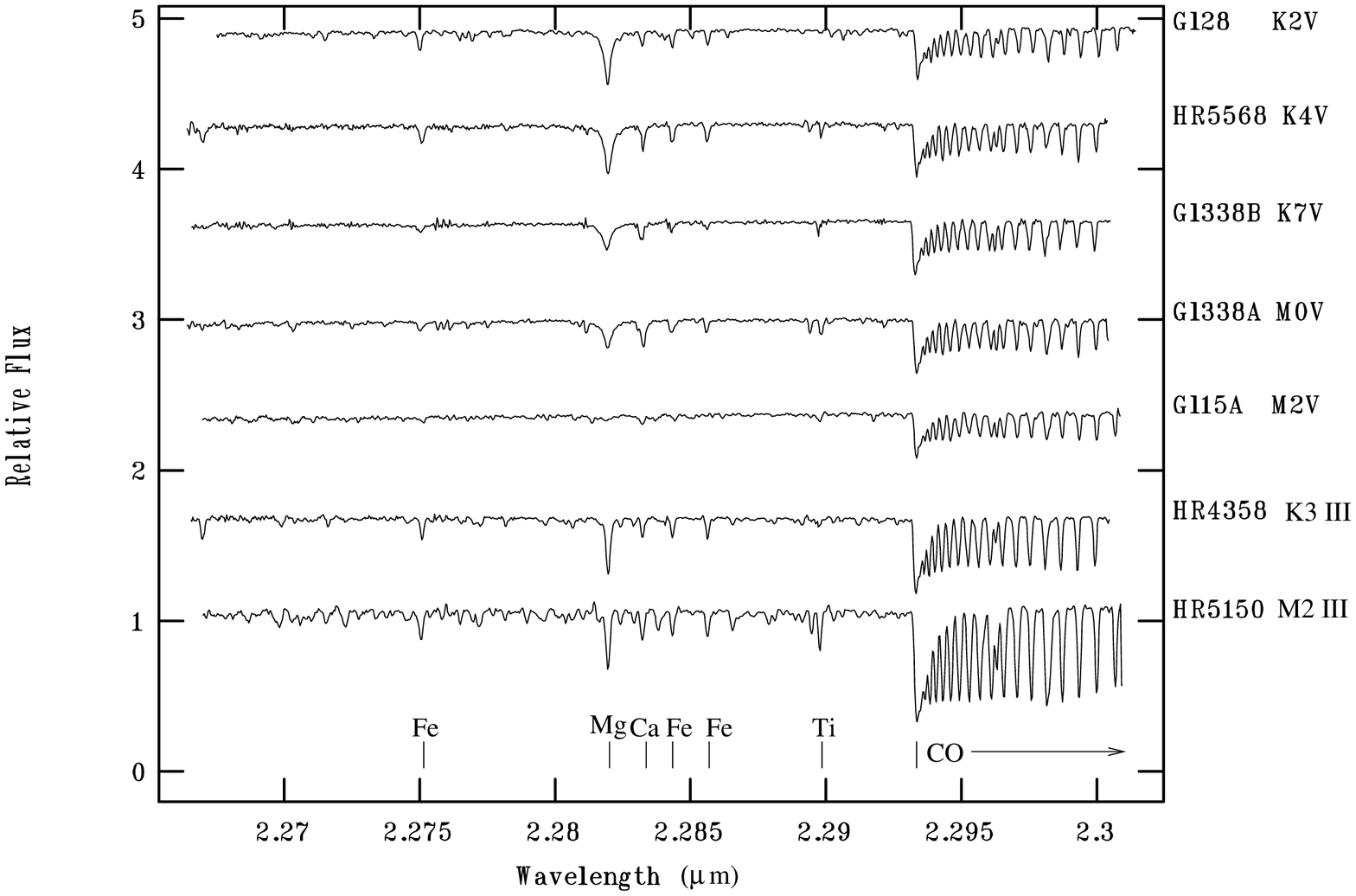}
\caption{Spectra of MK standards. Three NIRSPEC
orders are shown, and strong absorption features are identified in
each order. Each of the three orders contain a number of lines
whose ratios are
useful for evaluating the spectral types of protostars independent of
their veiling. Each spectrum has been normalized to a value of 1.0 and
they are offset vertically by a constant in each order for separation.}
\end{figure}

The two giant spectra (K3 III and M2 III) shown in Figure 1 have many
of the same lines as the dwarfs, but there are significant
differences. The Si and Mg lines in the shortest order do not appear
to change monotonically with spectral type, but the lines in the next
order are much more telling. The Sc (2.2058, 2.2071 $\mu$m) lines are
very strong (absolutely and relative to Na) in the M2III giant, while the Ti
lines are very strong in both giants. The Na lines (2.2062
and 2.2090 $\mu$m) are less broad in the giants because they are formed
under conditions of lower 
pressure than in dwarfs.  The Mg (2.2814 $\mu$m) line is also
less broad in the giants and does not vary
monotonically with spectral type. The 2.2935 $\mu$m CO band head and
nearby rotation-vibration lines are very strong
in the giants. Numerous other low-strength absorptions are seen in
all three orders of the giant star spectra, indicating the presence of
various molecules \citep[see also][]{WH96}.

Three NIRSPEC spectral orders of the YSOs WL 6 and YLW 15 are shown in 
Figure 2. The WL 6 data have signal-to-noise of 100 -- 170 at most
wavelengths, and the YLW 15 data have signal-to-noise 240 -- 300.
Both objects have strong continua rising towards longer wavelengths.
YLW 15 has numerous absorption features which are weak and moderately
broad, while any absorption features in the spectra of WL 6 are either
weaker or non-existent.
Both sources exhibit strong $v$ = 1 -- 0 S(0) H$_{2}$ emission
at 2.2233 $\mu$m.  Interestingly, the strength
of this emission appears to be correlated with infrared excess and
veiling in these objects.  WL 6 has weaker (absent) absorption lines, 
a stronger red continuum, and stronger H$_{2}$ emission than YLW 15.
 
\begin{figure}
\figurenum{2}
\plotone{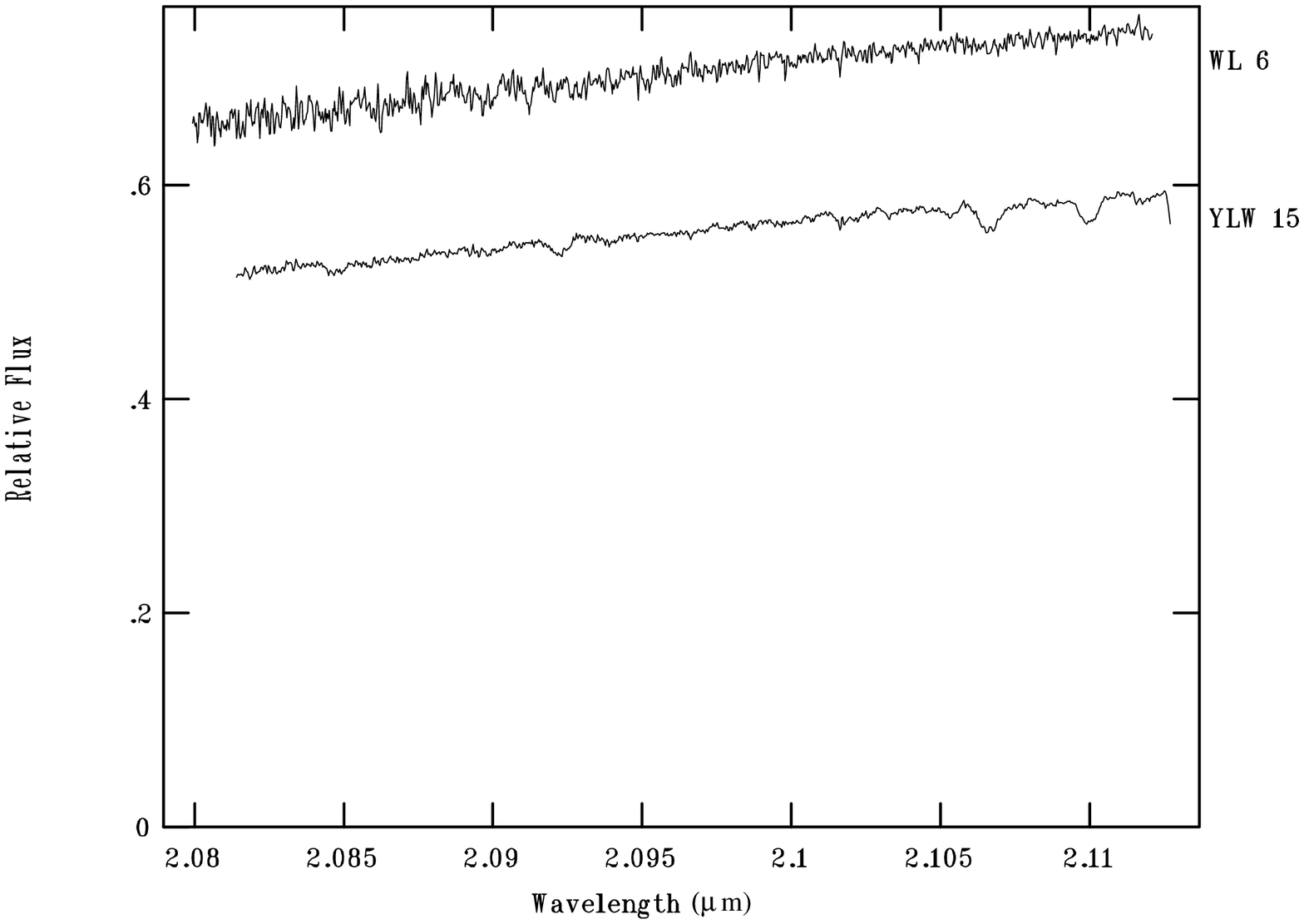}
\plotone{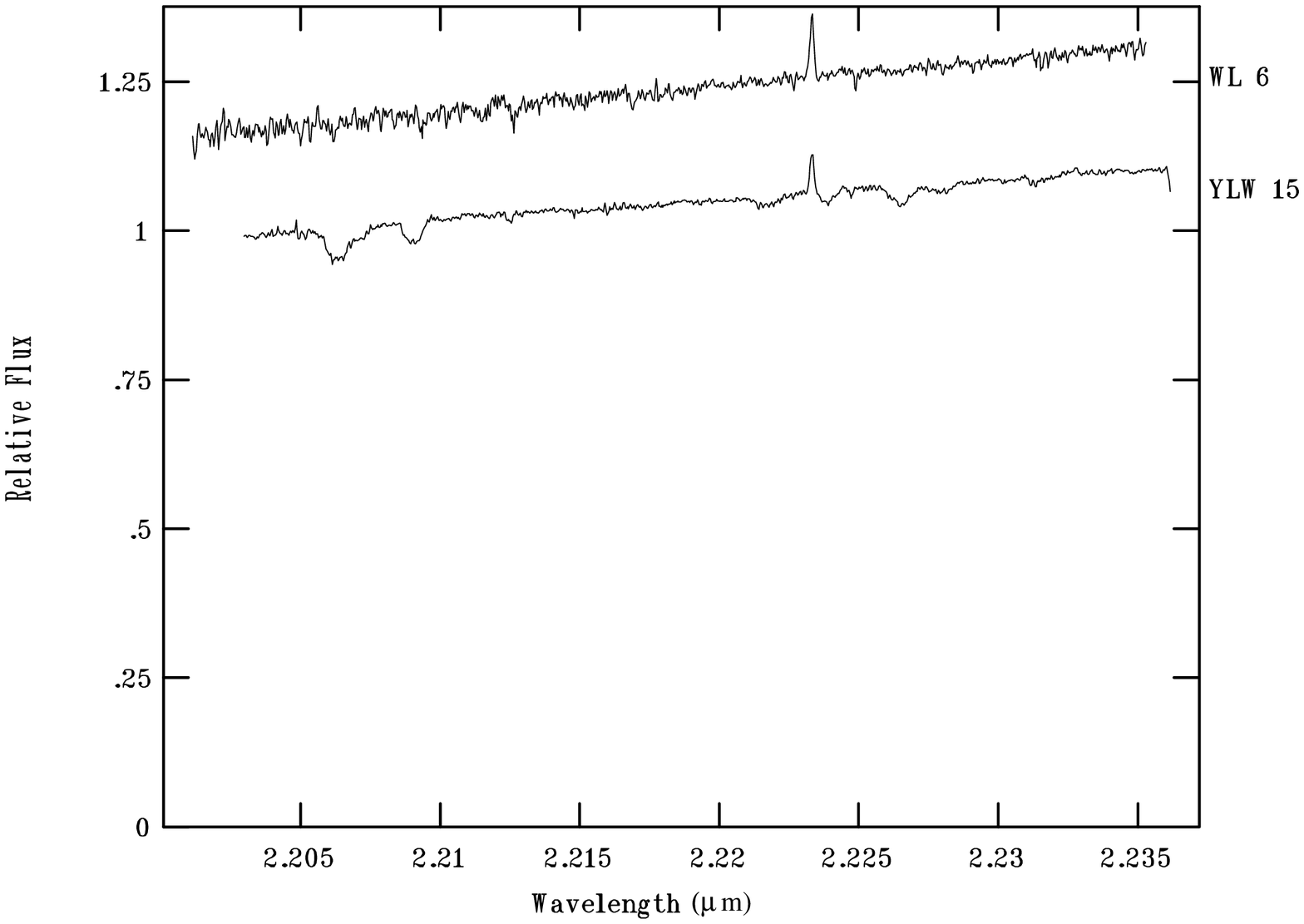}
\plotone{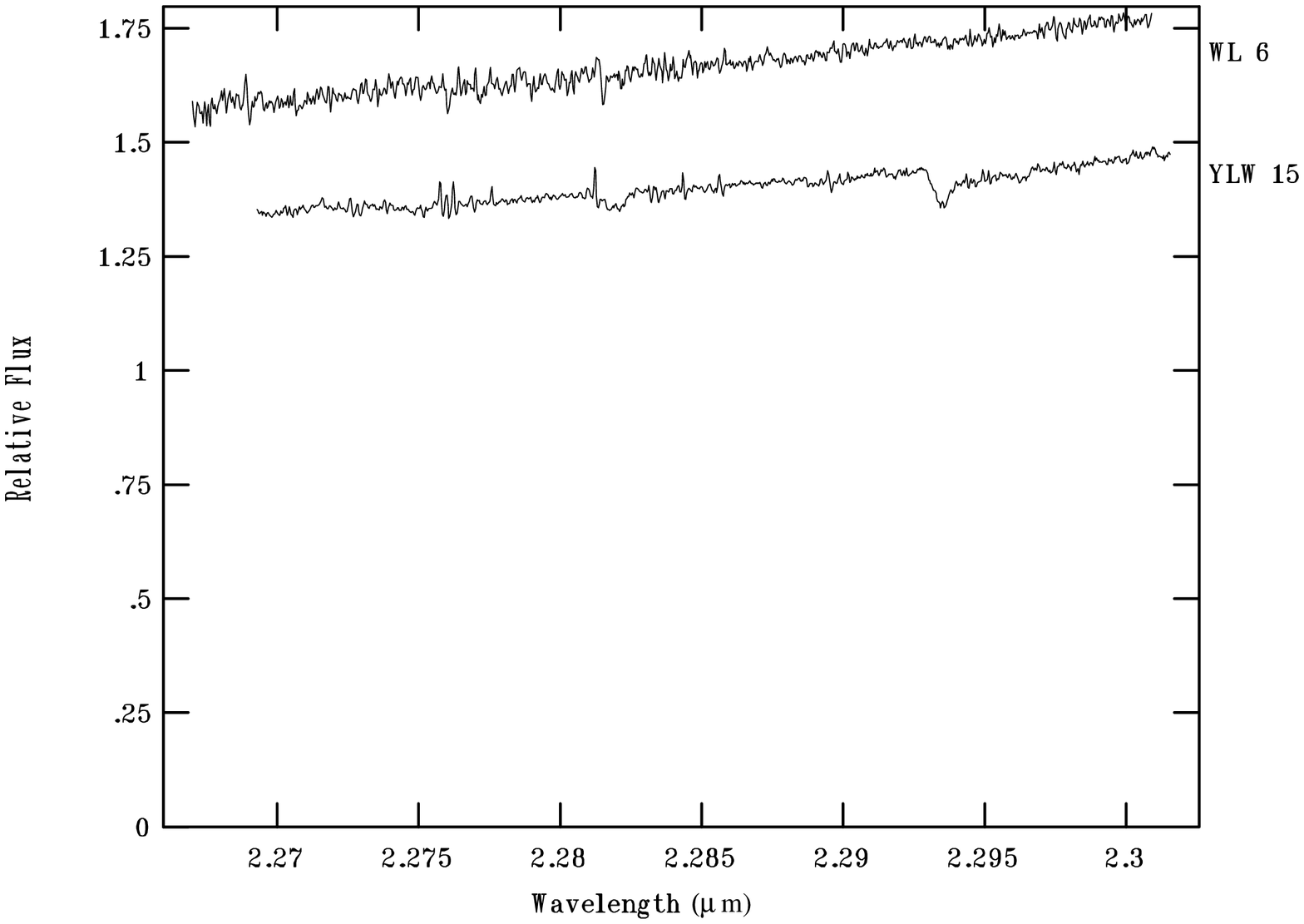}
\caption{Spectra of Class I protostars
YLW 15 and WL 6. The absorption lines are either weak (YLW 15) or
undetected (WL 6) due to the large continuum veilings ($r_k \geq 3$)
of these objects. Each spectrum has been normalized to a value of 1.0
(over all 3 orders), and WL 6 has been offset vertically by 0.2 for 
in each order for separation. }
\end{figure}

\subsection{Veiling and Rotation Analysis}

The spectra of YLW 15 and WL 6 are compared to those of the K4
dwarf HR 5568 in Figure 3. In this figure, the spectra of the YSOs are
processed beyond that shown in Figure 2 in order to facilitate
this comparison. First, their continuum slopes were removed
by normalization. Next, the relative strengths of their absorption lines
were magnified by subtracting off a constant continuum value of 0.75,
75\% of the normalized continuum. This specific constant was chosen
so that the absorption lines would have nearly the same equivalent
widths in both YLW 15 and HR 5568.  The
spectrum of HR 5568 was modified by convolving it with a limb-darkened
stellar rotation profile with velocity $v$ sin $i$ $=$ 50 km s$^{-1}$.
This was required to match the velocity widths of its absorption features
to those of YLW 15.

\begin{figure}
\figurenum{3}
\plotone{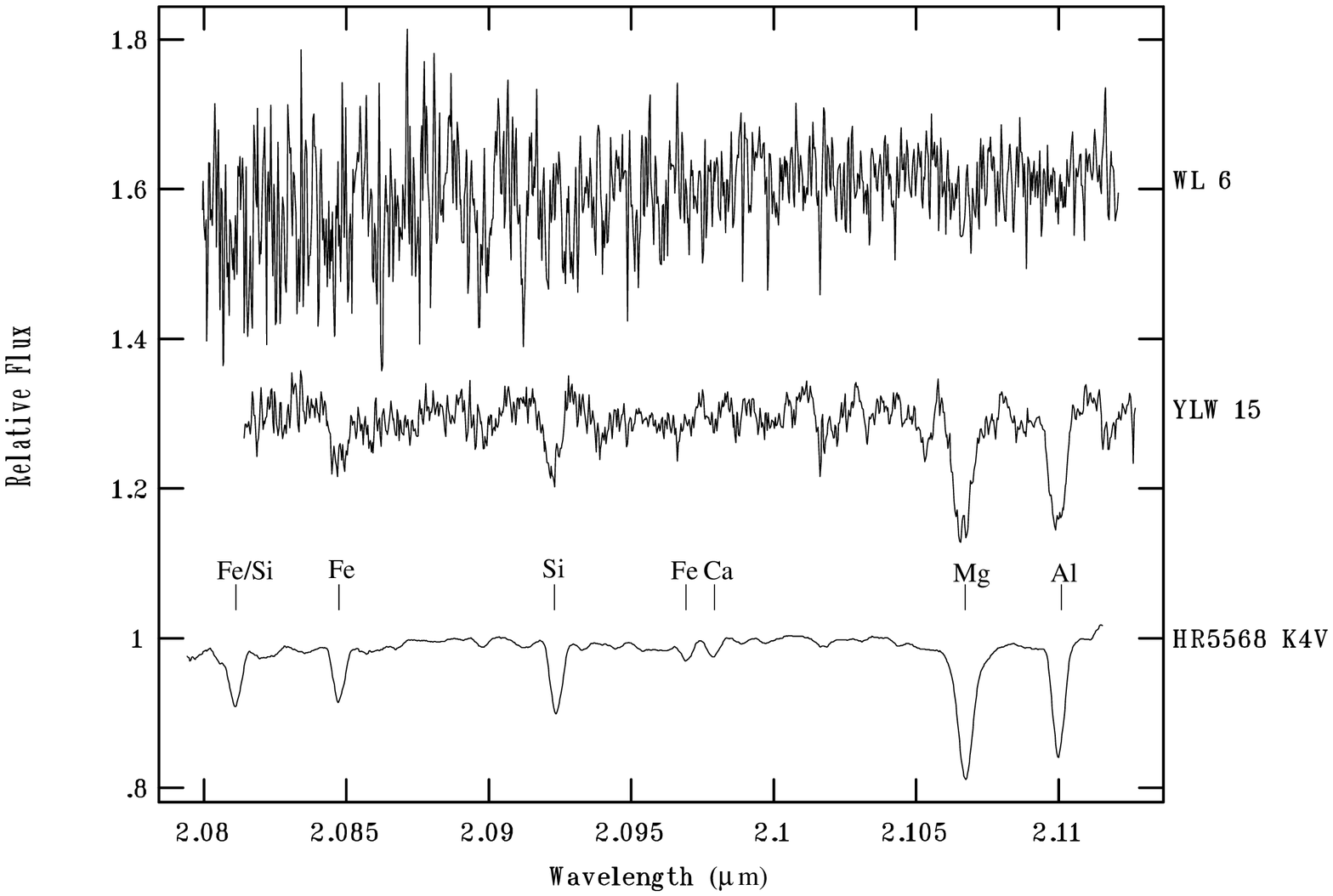}
\plotone{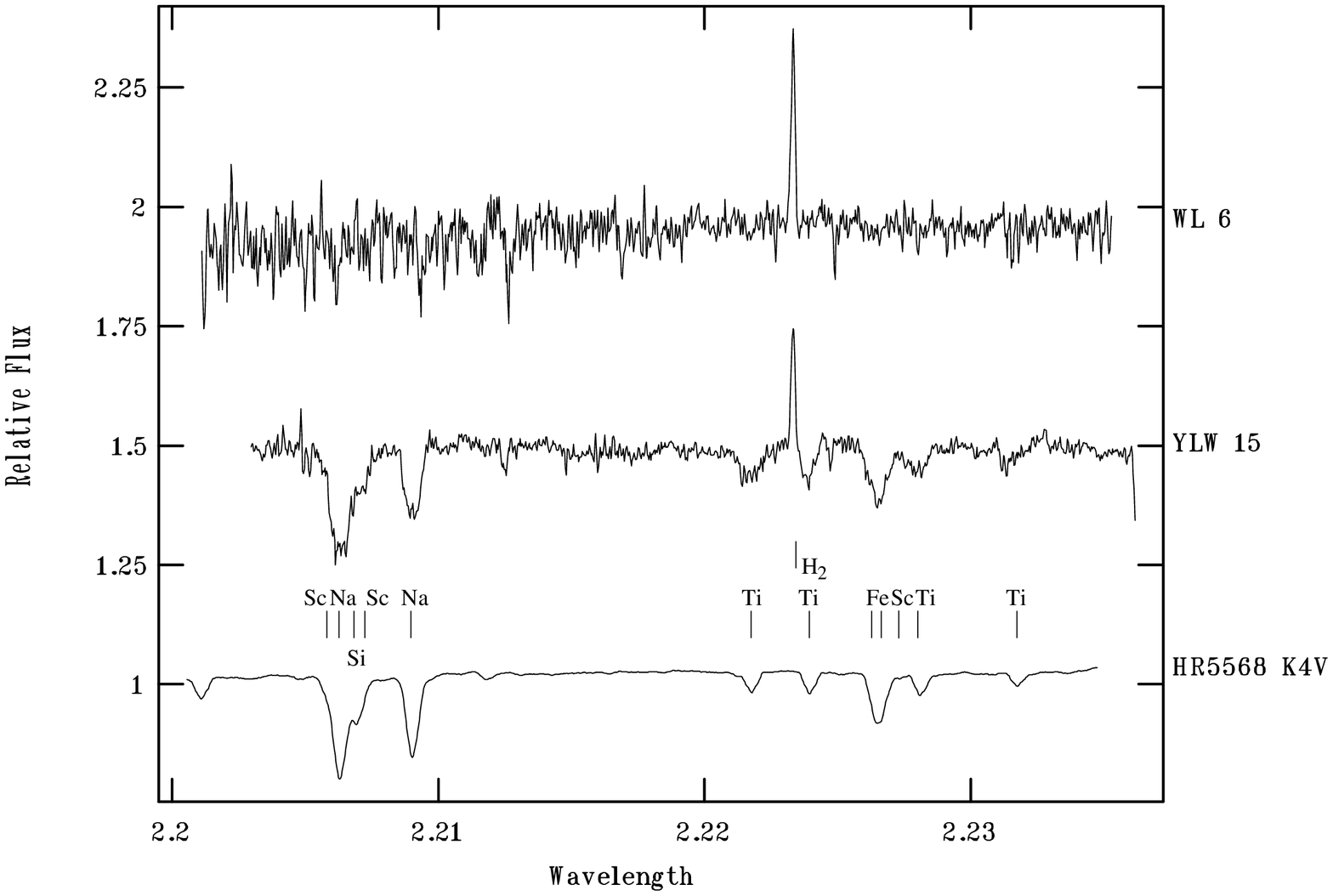}
\plotone{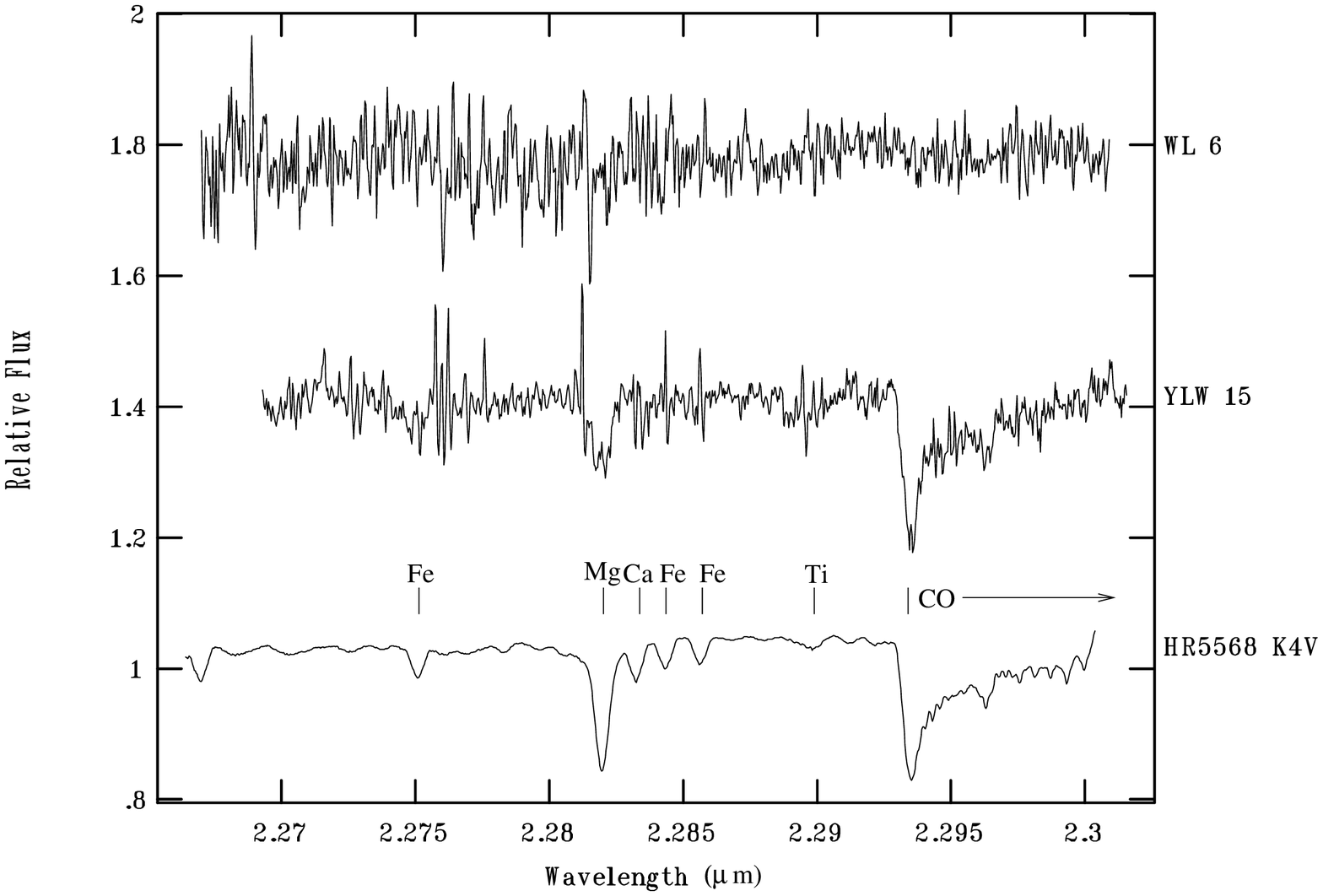}
\caption{Spectra of YLW 15 and WL 6 compared to HR
5568 (K4 V). The spectra of the protostars have been normalized to
remove
their continuum slopes, and 75\% of their continuum levels have been
subtracted to remove the continuum veiling amount $r_k = 3$. Each
spectrum has been normalized to a value of 1.0. The spectra are
offset vertically by a constant in each order for separation.
}
\end{figure}

Figure 3 shows that the spectrum of YLW 15 is very similar to that of
HR 5568. Nearly all absorption features have nearly identical
{\it relative} strengths. Comparing relative line strengths is most
useful because ratios (but not absolute equivalent widths) are
insensitive to continuum veiling. For example, the ratios of their Mg
(2.1066 $\mu$m) and Al (2.1099 $\mu$m) equivalent widths are very
similar, as are the ratios of their lines in the Na and Sc
region (2.2053 -- 2.2096
$\mu$m). The lines in the Na and Sc region also have the same
strengths relative to the Fe and Sc lines in the 2.2263 -- 2.2273
$\mu$m region. However, the spectra of the two objects do not match
exactly. The 2.2280 and 2.2317 $\mu$m lines of Ti are stronger
relative to the 2.2053 -- 2.2096 $\mu$m Na, Sc, and Si lines in YLW 15
than in HR 5568. Likewise, the 2.2814 $\mu$m Mg line is weaker relative
to the CO band head (or the 2.2053 -- 2.2096 $\mu$m Na, Sc, and Si
lines) in YLW 15 than in HR 5568. Both of these differences suggest
that YLW 15 is of slightly later spectral type than HR 5568. It is
clear from Figure 1 that YLW 15 is not as late as Gl 338B (K7 V) and
that its spectrum is much more dwarf-like than giant-like (see also
the discussion of giant spectra in \S 3.1).  We assign the
spectral type K5 IV/V to YLW 15. Differentiating between luminosity
class IV and V will probably require comparisons to synthetic model
spectra given the rarity of well-classified luminosity class IV
stars.  

The continuum veiling of YLW 15 can be measured now that its spectral
type has been determined. Continuum veiling at wavelength $\lambda$ is 
defined as the value of continuum excess expressed as a fraction of 
the stellar photospheric flux, $r_{\lambda}=F_{\lambda,ex}/F_{\lambda*}$. 
We have estimated the $K$--band continuum veiling of YLW 15, $r_k$,  by
comparing the equivalent width of its K-band spectral features to those of MK
standards. The above continuum veiling equation can be applied to
K-band wavelengths and written in terms of the intrinsic ($EW$) and 
observed ($EW^\prime$) equivalent widths of spectral features there,
$EW^\prime = EW / (1+{\rm r}_k)$.

We have evaluated $r_k$ of YLW 15 by comparing the measured equivalent
widths of its 2.2053 -- 2.2096 $\mu$m Na, Sc, and Si lines (from its 
spectrum in Figure 2) to those
of the K4 V star HR 5568 and the K7 V star Gl 338B. The measured
equivalent widths of these features are 0.8, 2.8, and 3.5 \AA~ for
similarly-reduced spectra of YLW 15, HR 5568, and Gl 338B,
respectively. The intrinsic equivalent widths of these features is
expected to be approximately 3.2 \AA~ for YLW 15, intermediate between HR
5568 and Gl 338B. Substituting these numbers into the above
equation yields $r_k$ $=$ 3.0. We estimate that this value is accurate
to about 25\% given the uncertainties in the measured equivalent
widths and the spectral type determination of YLW 15.
This veiling implies that 75\% of the $K$--band flux of YLW 15 arises
from (circumstellar) continuum excess, and 25\% of its flux arises
from its K5 stellar photosphere. Therefore the stellar spectrum of this
object can be approximated by subtracting 75\% of the continuum level
at $K$--band wavelengths. This -- and normalization by its continuum
-- was done to produce the spectrum of YLW 15 shown in Fig. 3.

The derived veiling $r_k = 3.0$ of YLW 15 is considerably less than
the $r_k \gtrsim 6$ which we estimated for this source in Paper II.
There may be several reasons for this discrepancy. First, we have
re-examined the older, lower quality 
data of YLW 15 and have concluded that CO band head absorption may be
present at low confidence in its spectrum (see Figure 2 of Paper II). 
The presence of such absorption
 would have lowered our estimate of the amount of veiling 
for this source. Second, the veiling in the protostar may
have decreased during the 4 years which elapsed between our two
observations. \cite{LR99} found that $\rho$ Oph sources can undergo
large variations in veiling, and they also derive $r_k > 1$ from their
2 $\mu$m spectrum of YLW 15. Finally, in Paper II we assumed that YLW
15 was spectral type M0 when we estimated its veiling. Its actual
spectral type is K5 (see above), which is expected to have
intrinsically weaker CO absorptions (analyzed in paper II) than 
would a M0 star. This would have led to an overestimate of its
veiling. 

We have determined that the projected rotation velocity of the YLW 15 central
protostar is $v$ sin $i = 50 \pm 5$ km s$^{-1}$. This was measured by
fitting its CO band head spectrum (Fig. 3c) to the spectrum of
HR 5568 (K4 V) which had been convolved with stellar rotation profiles
over a range of velocities. In order to confirm that this value was
not influenced by any slight rotation of HR 5568 (whose rotation is
unresolved in our spectra), we also fit the spectrum of YLW 15 with that
of a slowly-rotating K giant star which had been convolved with 
stellar rotation profiles over a range of velocities. 
This also resulted in a projected rotation velocity of 50 km s$^{-1}$.

The spectra of WL 6 (Figs. 2 and 3) do not show any obvious absorption
features, so only a lower limit to its veiling can be calculated. If
the central protostar of WL 6 has the same spectral type and $v$ sin
$i$ rotation velocity as YLW 15, then the signal-to-noise of its
spectrum ($\sim$ 120 in the 2.2053 -- 2.2096 $\mu$m region) dictates
that WL 6 must have $r_k \geq 4.6$.  This is consistent with the lower
limit determined in Paper II, $r_k > 3.8 - 5.5$. This veiling analysis
was based on maximum line depths (and not equivalent widths), so we
expect the veiling of this object to be
changed little if it is of later spectral type than YLW 15.
Our analysis of dwarf standards indicates that although equivalent
widths of the 2.2053 -- 2.2096 $\mu$m
features increase in later types (particularly in the M2 -- M6 range),
this is mostly caused by increasing intrinsic line widths (due to
pressure broadening), with the peak line depths relatively unchanged.
However, less veiling is
required if WL 6 is rotating more rapidly than YLW 15, and more is
required if it is rotating more slowly. Rotation can affect the
measured line depths and veilings by about a factor of 2 over the
range $v$ sin $i$ = 0 -- 180 km s$^{-1}$ (see Paper II).

\section{Discussion}

We have determined that the Class I YSOs WL 6 and YLW 15 have strong
continuum veilings, and both have near-IR H$_2$ emission. Absorption
lines in the spectrum of YLW 15 have allowed us to measure
quantitatively its spectral type, $r_k$ continuum veiling, and
$v$ sin $i$ rotation velocity. We now examine and discuss the physical 
implications of these results given other information known about
these YSOs.

\subsection{YLW 15}

The SED of YLW 15 rises steeply from the near- to the far-IR (2 -- 60
$\mu$m) with a bolometric luminosity of approximately 10 L$_\odot$
\citep{WLY89}. Its 2 -- 14 $\mu$m spectral index (slope of its SED)
is 1.0 \citep{SB01}. \citet{AM94} found that this source also has significant
$\lambda$ = 1.3 mm spatially-extended flux (100 mJy), which they
interpreted as arising from its circumstellar envelope, confirming its
Class I nature. \citet{BATC96} found that YLW 15 has a compact
molecular outflow which is seen nearly pole-on. The coexistence of
molecular outflow and near-IR H$_2$ emission (Figs. 2 and 3) in this
object is consistent with the correlation of these phenomena found by 
\citet{GL96}.  GY 263 is a fainter near-IR source \citep[$\delta K$ = 2.5
mag;][]{BKLT97} 6\arcsec~($\sim$ 900 AU projected distance) NW of YLW 15.

However, \citet{GRC00} recently found that YLW 15 itself is binary in their
$\lambda$ = 3.6 cm VLA map. 2MASS (second incremental release) 
Point Source Catalog coordinates show that the 2 $\mu$m source is essentially
coincident with VLA 2, the weaker, point-like radio source. VLA 1 is
located approximately 0\farcs6 ($\sim$ 100 AU) to the NW and is
spatially extended ($\sim$ 1\arcsec) SW -- NE. \citet{GRC00} interpret
the VLA 1 emission to be thermal free-free, arising in a protostellar
jet. HST / NICMOS near-IR images (F110W and F160W bands) show that the
central source is single but embedded in extended nebulosity
\citep{AMDMCY02}. Hereafter we assume that there is a central
protostellar engine for YLW 15, and it is located at the VLA 2 /
near-IR position. The VLA 1 source is likely to be emission from a jet
produced by the central protostar and not a (proto-) stellar source
itself.

One of the most intriguing aspects of YLW 15 is its X-ray emission.
This was one of the very few Class I protostars which was detected by
ROSAT \citep{CMFA95}. Subsequent ROSAT HRI \citep{GMFACG97}, ASCA
\citep{TIKGM00}, and Chandra \citep{IKT01} observations have all shown
that its X-ray emission has frequent strong flares \citep[$\sim
10^{33}$ erg s$^{-1}$; see][]{GMFACG97,IKT01}. The ASCA and Chandra
observations have shown that its quiescent and flaring emissions consist
of mostly hard X-rays. The ASCA observation
captured 3 quasi-periodic flares with a $\sim$ 20 h period which
spanned the exposure \citep{TIKGM00}, but the Chandra exposure
captured only a single flare ($\sim 3$ hr duration) in a 27.8 h exposure.
\citet{MGTK00} have
interpreted this flaring as arising from the interaction of the
magnetic field of the rapidly rotating protostar (P $\simeq$ 20 h, 
near break up) and its slower rotating circumstellar disk.

Our near-IR spectrum of YLW 15 provides further
information on the nature of this object. We have assigned
it a spectral type of K5 IV/V (see \S 3.2), implying an
effective temperature T$_{eff} \simeq 4300$ K \citep{Tok00}. 
Its near-IR spectrum (Fig. 3) shows no evidence for
binarity; it shows no features which could arise in a companion of
different spectral type. We now derive additional stellar and system
parameters for this object.

\subsubsection{Stellar Parameters}

We used new near-simultaneous near-IR photometry \citep[acquired
2001 July 8; see][]{HBGR02} to estimate the extinction towards the
central protostar of YLW 15. The observed CIT magnitudes of this object
were $H = 12.53$ and $K = 9.35$ (undetected at $J$), and we de-reddened the
$H-K$ color to determine that the extinction to the central protostar
is $A_v \simeq 40$ mag. This value is intermediate between the lower
limit ($A_v = 34$ mag), resulting from de-reddening to the maximum CTTS
locus color $(H-K)_0 = 1.0$ mag \citep[see][]{MCH97}, and the upper
limit ($A_v = 47$ mag), resulting from de-reddening to the intrinsic
stellar photospheric color $(H-K)_0 = 0.11$. YLW 15 cannot be accurately
de-reddened to this stellar photospheric color because the de-reddened colors 
of  Class I YSOs do not intersect stellar loci in near-IR color-color
diagrams. This is due to the presence of both scattered light and 
infrared excess emission from the close-in circumstellar material contained in 
the surrounding accretion disk and
infalling envelope \citep [see][] {LA92}.  Our adopted extinction
($A_v = 40$ mag) is consistent with the $A_v > 31$ mag value derived by
\citet{GMFACG97} using a similar technique. Our extinction value, along
with the measured near-IR excess $r_k = 3.0$, the IR extinction law,
dwarf bolometric corrections, and a distance modulus of 6.0 yields an
estimated {\it stellar} luminosity of 3.0 $L_\odot$. The 4300 K
effective temperature of the protostar dictates that its stellar
radius is 3.1 $R_\odot$. Although there are considerable uncertainties in 
extinction (and scattering), the derived protostellar properties are not
greatly affected.  Varying the $A_v$ between its lower and upper
limits changes the derived stellar luminosity by less than a
factor of 2 and the derived stellar radius by less than 50\%. The
derived luminosity and radius are very similar to those predicted for
a young PMS star of this effective temperature which is still accreting mass 
at a high rate \citep[$\dot{M} \simeq 10^{-6} M_\odot$ yr$^{-1}$;
see][]{SFB99, TLB99}. These and other modern PMS models indicate that 
the mass of YLW 15 is approximately 0.5 M$_\odot$, as determined from
its effective temperature and stellar luminosity \citep{DM97, SFB99, 
SDF00}.  Taken with its derived radius, this dictates that the surface
gravity of YLW 15 is Log $g \simeq$ 3.1 (cgs units), within the range
expected for a very young K5 PMS star. 

\subsubsection{Accretion}

The derived stellar luminosity is about 30\% of YLW 15's {\it bolometric}
luminosity \citep{WLY89}, implying that 70\% of the total luminosity of this
protostar is due to mass accretion and not radiation of the protostar's
internal energy. 
This accretion luminosity is likely to be composed of a component which 
arises in a circumstellar accretion disk as well as a component which
is generated when accreting matter falls directly onto the central
protostar. If all matter accretes through a circumstellar disk with
inner radius $R_{in}$, then the accretion luminosity arising in the 
disk as $GM\dot{M_D}/2R_{in}$ where $\dot{M_D}$ is the mass accretion rate
through the disk. The accretion luminosity produced by matter 
falling from the disk edge $R_{in}$ onto the star is $GM\dot{M_*}/R_* 
- GM\dot{M_*}/R_{in}$ where $\dot{M_*}$ is the mass accretion rate onto
the star. We now assume that
matter is transported from near the disk edge to the stellar
surface in magnetospheric accretion columns \citep{K91, SNOWRL94}.
In this scenario, matter accretes through the disk to a disk coupling radius
$R_x$ where it is then accreted onto the star along magnetic field
lines \citep{SNOWRL94}. This point $R_x$ is slightly exterior to $R_{in}$, and 
the total accretion luminosity of the system is

\begin{equation}
L_{acc} = {G M_* \dot{M_D}\over 2R_x} + {G M_* \dot{M_*}\over R_*} 
- {G M_* \dot{M_*} \over R_x}.
\end{equation}

The typical inner hole radius of a CTTS circumstellar disk
is $R_{in}\simeq 5 R_*$ \citep{SNOWRL94, MCH97}.
Magnetospheric accretion models predict that the inner disk and
coupling radii are
closer to the central star when matter is accreting at higher rates, 
such as those typical for Class I protostars. DQ Her stars are also
sometimes interpreted as exhibiting this behavior \citep[e.g., see][]{P94}.
We assume $\dot{M_*} = 0.7 \dot{M_D}$ from \citet{SNOWRL94}
and adopt their equation (2.6b) to express the magnetic coupling radius as

\begin{equation}
{R_x \over R_*} = \alpha_x \left({{B_*^4 R_*^5} \over
	{GM_*{\dot{M}_D}^2}} \right)^{1/7}
\end{equation} 

where $\alpha_x$ is a constant on the order of 1 \citep{SNOWRL94,
OS95}.

We solved equations (1) and (2) simultaneously for mass accretion
rates and $R_x$ using the previously derived values of $R_*$ (3.1
$R_\odot$), $M_*$ (0.5 $M_\odot$), $\alpha_x = 0.92$ \citep{OS95}, 
the above relation of mass accretion rates, and the assumption
$B_* = 1000$ G. This results in
a magnetic coupling radius $R_x = 2.1 R_*$ and mass accretion rates
$\dot{M_D} = 2.3 \times 10^{-6} M_\odot$ yr$^{-1}$ and $\dot{M_*} = 1.6
\times 10^{-6} M_\odot$ yr$^{-1}$. This magnetic coupling radius is
considerably uncertain (perhaps by 50\%) because we have assumed a
magnetic field strength and have not measured a value. The 2.2233
$\mu$m TI line is sensitive to magnetic (Zeeman) broadening, and
indeed it has a larger equivalent width ($\sim 0.4$ \AA) in the 
stellar spectrum of YLW 15 than in either the K4 V standard HR 5568
(0.2 \AA) or the K7 V standard Gl 338B (0.3 \AA; see Figs. 1b 
and 3b). This suggests the presence of a significant \citep[kG; e.g.
see][]{JVK99} magnetic field in YLW 15, but the magnitude of this
field cannot be measured quantitatively from our spectrum due to its
limited signal-to-noise and the presence of the flanking 2.2233 $\mu$m
H$_2$ line (Fig. 3b).

The derived accretion rates are consistent with the accretion 
rates computed for other Class I YSOs in this cloud
and are within a factor of 2 of the spherical accretion rate derived
from the cloud's gas temperature \citep[$\dot{M}=a^3/G$;
see][]{ALS87}. The adopted 0.5 $M_\odot$ stellar mass of YLW 15 is
also consistent with accretion at this rate (i.e. $\dot{M_*}$) over
the mean Class I lifetime in the cloud \citep{GWAYL94,LR99}.
The derived accretion rates and
inner disk radius result in accretion luminosities of 2.75 $L_\odot$
arising in the disk and 4.25 $L_\odot$ arising at the stellar surface.
Thus the stellar surface accretion luminosity is approximately 40\% greater
than the luminosity emanating from the $T_{eff} = 4300$ K protostellar
photosphere.
\citet{CG98} have shown that the region where an accretion column
meets the surface of a T Tauri star has an effective temperature
of approximately 8000 K. From this we calculate that the derived 
4.25 $L_\odot$ stellar surface accretion luminosity arises from an
area only 12\% of the stellar surface.

\subsubsection{Veiling}

Here we investigate the possible sources of excess infrared
continuum emission and their contributions
to the overall veiling in the infrared spectrum of YLW 15. 
We consider three sources for the continuum veiling: the accretion
luminosity generated at the surface of the protostar, the 
luminous accretion disk, and the inner infalling envelope.
We calculate that the continuum from the hot accretion impact sites at the 
surface of the protostar will produce an
effective veiling $r_k = 0.33$. Although a relatively small 
fraction of the total veiling, this is significantly greater
than that predicted ($r_k < 0.05$) for accretion impact
sites on typical class II T Tauri stars whose accretion rates are
substantially lower \citep{CG98}.  Accretion models predict that the
disk of YLW 15 ($L_{disk} = 2.75 L_\odot$) should contribute an
IR veiling $r_k \sim 1.3$ \citep[see Fig. 9 of][]{GL96}.
The remainder
of the object's observed veiling ($r_k \sim 1.4$ of $r_k = 3$ total)
is likely to arise from  grains in
the circumstellar envelope of the protostar which are warmed by the
photons produced by the stellar photosphere and at the hot sites of mass
accretion. This implies that the envelope contains a significant 
amount of dust within $\sim$ 0.3 AU of the central star \citep[see
also][]{GL96, CHS97}. 

We note that the envelope veiling is low enough so
that any absorption lines originating in the circumstellar disk should
have been seen in the spectrum of YLW 15. However, the (stellar)
dwarf-like natures of the YLW 15 absorption features imply that they
are not formed in a circumstellar disk, which would show evidence of
much lower surface gravity (see Paper I for a discussion and
examples of FU Ori disk features). 
We calculate that the dust grain destruction distance 
(where $T \lesssim 1800$ K) for YLW 15 is approximately 10 $R_*$
from considering both its protostellar photospheric and 
stellar accretion luminosities. This implies
that the disk is completely gaseous and dust-free between $R_{in} = 2
R_*$ and 10 $R_*$. The fact that we do not observe absorption lines
arising in this disk region strongly implies that the gas in this region
does not have a sufficient vertical temperature gradient to form
strong spectral lines. 
 
\subsubsection{Rotation}

The $v$ sin $i$ = 50 km s$^{-1}$ projected rotation velocity of YLW 15 is
considerably greater than typical for many classical T Tauri stars, about
equal to that found for flat-spectrum protostars in the $\rho$ Oph cloud 
(see Paper I), and considerably less than the equatorial rotation velocity
deduced for this object from its X-ray emission. If the 20 h X-ray flare
period observed by \cite{MGTK00} was caused by rotation as they
postulated, then assuming a 3.1 $R_\odot$ protostellar radius would 
imply an equatorial rotation velocity of approximately 180 km s$^{-1}$,
approximately break up. If YLW 15 is rotating this fast, then its rotation axis
must be only $16^\circ$ from our line of sight. This is consistent with its
molecular outflow being oriented nearly pole-on \citep{BATC96} if its 
outflow axis is aligned with its rotation axis. However, subsequent
Chandra X-ray observations of YLW 15 do not support the interpretation
that it is rotating with a 20 h period. These observations did not
confirm flaring with a 20 h period; only one flare was detected in a
$10^5$ s (27.8 h) observation \citep{IKT01}. This weakens the possibility
that the ASCA X-Ray flare period is physically related to the
protostellar rotation rate. The inclination -- and hence the
equatorial rotation velocity -- of YLW 15 is not well constrained.
In order to make quantitative comparisons, we assume that the rotation axis
of YLW 15 is inclined $57^\circ$ from our line of sight, the most likely
inclination angle for a random distribution. This implies that its
equatorial rotation velocity is approximately 60 km s$^{-1}$, but this
is highly uncertain: our newly-determined parameters and previous data do not
constrain this velocity well within the range $v_{rot}$ = 50 -- 180 km s$^{-1}$.

We estimate that the Keplerian co-rotation radius of the YLW 15 circumstellar
disk is 2.0 $R_*$ (for $v_{rot}$ = 60 km s$^{-1}$), nearly equal to
the $R=2.1 R_*$ magnetic coupling radius. The co-rotation radius varies
over the range of 2.3 -- 1.0 $R_*$ for the range of equatorial rotation
velocities $v_{rot}$ = 50 -- 170 km s$^{-1}$. Due to its large
uncertainty (see \S 4.1.2), we cannot determine whether the magnetic coupling
radius is significantly different from the co-rotation radius over a large range
of inclination angles. However, it appears that inclination
angles $i \sim 65\arcdeg$ (similar to our $57\arcdeg$ assumption) are most
compatible with the
\citet{SNOWRL94} requirement that the co-rotation and magnetic coupling
radii be essentially equal (provided that we have adopted the correct
value of $B_*$).

If the central protostar and the disk are sufficiently magnetically coupled
in this region, then the disk may be regulating the angular momentum 
(and rotation velocity) of the star \citep[e.g.,][]{K91,Eetal93,SNOWRL94}.
Although the observational evidence of such coupling in CTTSs has been recently
challenged \citep[e.g., see][]{SMMV99,SMVMH01}, 
the similarity of the co-rotation and coupling radii of YLW 15 are
consistent with stellar spin-down and magnetic star--disk coupling.
Indeed, we calculate that the magnetic braking time scale, $\tau \gtrsim
\vert \dot{J_*} / (dJ/dt) \vert$ 
\citep[F. Shu priv. communication;
eq. 6 of][]{H02}, is only $\gtrsim$ 14,000 yr for YLW 15. This is
significantly shorter than the $\sim 10^5$ yr age of Class I
protostars, so YLW 15 is likely to have had more than enough time to couple
magnetically to its disk.

Class II YSOs and CTTSs in the $\rho$ Oph
cloud \citep [and CTTSs in Taurus; see][]{Eetal93} are observed to rotate
considerably more slowly ($v$ sin $i \simeq 12$ km s$^{-1}$;
see Paper I), so if YLW 15 is to evolve into a typical Ophiuchi CTTS
it will likely lose about 75\% of its angular momentum. This can
result via disk braking if the protostar remains coupled or recouples
to its disk as its mass accretion rate drops.
If the mass accretion rate of YLW 15 suddenly
dropped to $10^{-7} M_\odot$ yr$^{-1}$, then it would re-couple to its
disk in only about 150,000 yr \citep[i.e. eq. 6 of][]{H02}. This
coupling would occur at a radius $R \simeq 4.5 R_*$ with 
co-rotation speed $v$ sin $i \simeq 15$ km s$^{-1}$ (for $i = 57^\circ$),
very similar to the mean rotation velocity observed for $\rho$ Oph Class
II YSOs. We note that this 150,000 year coupling time is much shorter
than the mean age of $\rho$ Oph Class II YSOs (5 -- 10 $\times 10^5$
yr), so this can explain why Class II YSOs are observed to rotate more
slowly than Class I and flat-spectrum ones.
However, more sophisticated modeling is required to ascertain the
exact nature of this spin-down behavior.

\subsection{WL 6}

\cite{SB01} found that WL 6 has a spectral index of 0.6 (over $\lambda$
= 2 -- 14 $\mu$m), less than that
of YLW 15. \cite{WLY89} also found that the WL 6 spectral index is less
than YLW 15, but their absolute values and the spanned wavelength range are
different. WL 6 has a bolometric luminosity of approximately 2 L$_\odot$
\citep{WLY89, SB01}. Neither \cite{AM94} or \cite{MAN98} detected WL 6
at $\lambda$ = 1.3 mm (peak $F_\nu < 20$ mJy in 11\arcsec~diameter),
but both investigations regarded it as an
embedded protostar due to its large spectral index. This source is not
known to drive a molecular outflow \citep{ BATC96}, but it does have
near-IR H$_2$ emission (Figs. 2 and 3). WL 6 source has no
known companions, and it also appeared single at $\lambda = 2$ $\mu$m
in the NIRSPEC camera images.

\cite{KKTY97} found that WL 6 had a hard X-ray spectrum and its flux
varied sinusoidally with a 23 h period, which they attributed to 
protostellar rotation. \cite{MGTK00} also modeled this variation as
being due to rotation, but with a slower period (3.3 d). \cite{IKT01}
did not detect this sinusoidal
variation, but they did detect two weak flares during their $10^5$ s
Chandra exposure. Taken with the results of YLW 15, we interpret these
differences to indicate that the X-ray variability of WL 6 is not
directly related to its protostellar rotation period. However, this cannot be
confirmed until the rotation signature of a photospheric feature (such
as 2.29 $\mu$m CO absorption) is observed in this object.

We have used the bolometric luminosity of WL 6 and the accreting PMS
models of \citet{SFB99} to bound the stellar parameters 
of this object. The stellar luminosity of WL 6 must be less than or
equal to its bolometric luminosity, 2 $L_\odot$. The \citet{SFB99}
models predict that such a PMS star at a very young age ($\sim 10^5$ yr)
must have mass $M_* \lesssim 0.3 M_\odot$ and $T_{eff} \lesssim 4000$ K
(their Case B). The high veiling of WL 6 (see \S 3.2)
implies that its stellar luminosity is considerably lower than 2
$L_\odot$. We now assume that WL 6 has approximately the same ratio 
of stellar to bolometric luminosity of YLW 15 since the $r_k$ of WL 6 is
somewhat larger and its $T_{eff}$ is somewhat lower than YLW 15, with
these effects tending to cancel each other. This implies that the
stellar luminosity of WL 6 is $L_* \simeq 0.6 L_\odot$, with
corresponding $M_* \sim 0.1 M_\odot$ and $T_{eff} \lesssim 3160$ K.
This temperature is consistent with that of a M3 -- M4 PMS star.  
Our previous models show that it is possible to produce significant
veiling from only a circumstellar accretion disk for a PMS star
of this mass \citep[see Fig. 9 of][]{GL96}. However, it is also 
plausible and possible that matter may also accrete directly from the
disk to the star, and some veiling could also be produced by its
circumstellar envelope (all as in YLW 15). These limits on the mass,
effective temperature, and luminosity of WL 6 imply that its mass accretion
rate is on the order of $10^{-6} M_\odot$ yr$^{-1}$, from the relation
$L_{acc} = L_{bol} - L_* = GM_*\dot{M}/R_*$. This rate is
consistent with the typical $\sim 10^5$ yr Class I lifetime and the
$\sim 0.1 M_\odot$ mass of WL 6.

\section{Summary and Conclusions}

We have presented the first spectrum of a highly veiled, strongly
accreting Class I protostar (YLW 15 / IRS 43) which shows late-type
photospheric absorption features. This spectrum has allowed us
to diagnose a plethora of details about the stellar, accretion, and
rotation properties of this protostar and its circumstellar
environment. The spectrum of another Class I protostar, WL 6,
did not show any photospheric absorption features. Both of these
objects are known to have strong and variable hard X-ray emission.
Our analysis of these spectra results in the following conclusions:

1. We have detected numerous late-type absorption features in the
photosphere of YLW 15. These features include Si (2.0923 $\mu$m and
2.2069 $\mu$m), Mg
(2.1066 $\mu$m), Al (2.1099 $\mu$m), Na (2.2062, 2.2090 $\mu$m), Sc
(2.2058, 2.2071 $\mu$m), Ti (primarily 2.2217 and 2.2240 $\mu$m),
and Mg (2.2814 $\mu$m) lines as well as the CO $v = 0 - 2$ band.
All of these features are consistent with a single K5 spectral
type ($T_{eff} = 4300$ K) with dwarf-like surface gravity (luminosity
class IV/V).

2.  YLW 15 has a low-mass central protostar ($M \simeq 0.5 M_\odot$)
with effective temperature, surface gravity, and {\it stellar} luminosity
similar to that of a T Tauri star despite its high bolometric luminosity
($L_{bol} \simeq 10$ $L_\odot$).  We find that the stellar luminosity
of YLW 15 is 3 $L_\odot$ and its accretion luminosity is 7 $L_\odot$.
Assuming the magnetic accretion model of \citet{SNOWRL94}, we derive
the disk mass accretion rate of YLW 15 to be $\dot{M_D}= 2.3 \times
10^{-6} M_\odot$ yr$^{-1}$ and its stellar mass accretion rate to be
$\dot{M_*}= 1.6 \times 10^{-6} M_\odot$ yr$^{-1}$. These rates are consistent
with the mass accretion rate of the ambient cloud gas and the stellar mass
of YLW 15 given the lifetimes of Class I protostars.  

3. YLW 15 is highly veiled in the near-IR, $r_k \geq 3$.  We explore
the origin of this veiling and evaluate the contributions of the
stellar surface accretion, circumstellar disk, and circumstellar envelope
components. We estimate that the circumstellar disk and envelope 
contribute equally to the veiling each providing $\sim$ 45\% of
the needed excess emission. We find a significant amount ($r_k \sim 0.33$)
of veiling is provided by the accretion shocks at the stellar surface.
Although this contribution is relatively small ($\sim$ 11\%) compared
to the total, it is significantly greater than that produced
by similar processes in typical CTTSs. 
We determine that emission from the circumstellar envelope
is inadequate to veil absorption features formed in the inner
dust-free circumstellar disk. We conclude that there are no strong
absorption lines formed in the warm inner disk of YLW 15.

4. The projected rotation velocity of YLW 15 is $v$ sin $i$ = 50 km s$^{-1}$,
much faster than that of a typical T Tauri star in the $\rho$ Oph
cloud. However, this rotation rate does not appear to be correlated
with the X-ray flarings and variability observed in multi-epoch
X-ray observations of this source.  We do find that it is plausible
that YLW 15 is magnetically coupled to its accretion
disk at a radius of 2 $R_*$ in the magnetic accretion model of
\citet{SNOWRL94}. This suggests -- and we verify the
plausibility -- that stellar angular momentum decreases by over a
factor of 2 in only $\sim 10^5$ yr between the protostellar and 
T Tauri evolutionary phases.

5. The near-IR spectrum of WL 6 does not show any obvious absorption
features, and we estimate that the veiling of this YSO is $r_k > 4.6$.
This high veiling and its low luminosity ($L_{bol} \simeq 2 L_\odot$)
dictate that its central protostar is likely to have low stellar
luminosity ($L_* \simeq 0.6 L_\odot$) and to be low mass ($M \sim
0.1 M_\odot$). 

\acknowledgments

We thank Scott Kenyon and Robbins Bell for useful comments and discussions.
The authors wish to recognize and acknowledge the very significant
cultural role and reverence that the summit of Mauna Kea has always had
within the indigenous Hawaiian community.  We are most fortunate to have
the opportunity to conduct observations from this mountain. 
We also thank the staff of the Keck
Observatory for excellent support in the operation of the telescope and
NIRSPEC.  This research has made use of data products from the Two Micron
All Sky Survey, which is a joint project of the University of
Massachusetts and the Infrared Processing and Analysis
Center/California Institute of Technology, funded by the National
Aeronautics and Space Administration and the National Science Foundation.
T. P. G. acknowledges grant support from the NASA
Origins of Solar Systems Program, NASA RTOP 344-37-22-11. Finally, we
thank NASA for providing and awarding time for these observations.


\begin{thebibliography}{}

\bibitem[Adams, Lada, \& Shu(1987)]{ALS87} Adams, F.~C., 
Lada, C.~J., \& Shu, F.~H.\ 1987, \apj, 312, 788 
    
\bibitem[Allen et al.(2002)]{AMDMCY02} Allen, L.~E., Myers, P.~C.,
Di Francesco, J., Mathieu, R., Chen, H., \& Young, E.\ 2002, \apj, 566, 993

\bibitem[Andr\'{e} \& Montmerle(1994)] {AM94} Andr\'{e}, P. \& Montmerle, T.
1994, \apj, 420, 837

\bibitem[Barsony et al.(1997)]{BKLT97} Barsony, M., Kenyon, S.~J., Lada,
E.~A., \& Teuben, P.~J.\ 1997, \apjs, 112, 109

\bibitem[Bontemps et al.(1996)]{BATC96} 
Bontemps, S., Andre, P., Terebey, S., \& Cabrit, S.\ 1996, \aap, 311, 858 

\bibitem[Bontemps et al.(2001)]{SB01} Bontemps, S.~et al. 2001, \aap, 372, 173

\bibitem[Calvet, Hartmann, \& Strom(1997)]{CHS97} Calvet, N., 
Hartmann, L., \& Strom, S.~E.\ 1997, \apj, 481, 912

\bibitem[Calvet \& Gullbring(1998)]{CG98} Calvet, N.~\& 
Gullbring, E.\ 1998, \apj, 509, 802 

\bibitem[Casanova et al.(1995)]{CMFA95} Casanova, S., Montmerle, T.,
Feigelson, E.~D., \& Andre, P.\ 1995, \apj, 439, 752

\bibitem[D'Antona \& Mazzitelli(1997)]{DM97} D'Antona, F.~\& 
Mazzitelli, I.\ 1997, Memorie della Societa Astronomica Italiana, 68, 807 

\bibitem[Edwards et al.(1993)]{Eetal93} Edwards, S.~et al.\ 
1993, \aj, 106, 372

\bibitem[Girart, Rodr{\' i}guez, \& Curiel(2000)]{GRC00} Girart, J.,
Rodr{\' i}guez, L.~F., \& Curiel, S.\ 2000, \apjl, 544, L153

\bibitem[Greene \& Lada(1996)]{GL96} Greene, T.~P.~\& Lada, 
C.~J.\ 1996, \aj, 112, 2184

\bibitem[Greene \& Lada(1997)]{PaperI} Greene, T. P., \& Lada, C. J. 1997, 
\aj, 114, 2157 (Paper I)

\bibitem[Greene \& Lada(2000)]{PaperII} Greene, T. P., \& Lada, C. J. 2000, 
\aj, 120, 430 (Paper II)

\bibitem[Greene et al.(1994)]{GWAYL94} Greene, T.~P., Wilking, 
B.~A., Andre, P., Young, E.~T., \& Lada, C.~J.\ 1994, \apj, 434, 614

\bibitem[Grosso et al.(1997)]{GMFACG97} Grosso, N., Montmerle, T.,
Feigelson, E.~D., Andre, P., Casanova, S., \& Gregorio-Hetem, J.\ 1997,
\nat, 387, 56

\bibitem[Haisch et al.(2002)]{HBGR02} Haisch, K. E. Jr.,
Barsony, M., Greene, T. P., \& Ressler, M. E., preprint

\bibitem[Hartmann(2002)]{H02} Hartmann, L.\ 2002, \apjl, 566, L29 

\bibitem[Imanishi, Koyama, \& Tsuboi(2001)]{IKT01} Imanishi, K., Koyama,
K., \& Tsuboi, Y.\ 2001, \apj, 557, 747

\bibitem[Johns-Krull, Valenti, \& Koresko(1999)]{JVK99} 
Johns-Krull, C.~M., Valenti, J.~A., \& Koresko, C.\ 1999, \apj, 516, 900 

\bibitem[Kamata et al.(1997)]{KKTY97} 
Kamata, Y., Koyama, K., Tsuboi, Y., \& Yamauchi, S.\ 1997, \pasj, 49, 461 

\bibitem[Kenyon et al.(1998)]{KBTB98} 
Kenyon, S.~J., Brown, D.~I., Tout, C.~A., \& Berlind, P.\ 1998, \aj, 115, 
2491

\bibitem[Koenigl(1991)]{K91} Koenigl, A.\ 1991, \apjl, 370, L39. 

\bibitem[Lada \& Adams(1992)]{LA92} Lada, C.~J.~\& Adams, 
F.~C.\ 1992, \apj, 393, 278

\bibitem[Luhman \& Rieke(1999)]{LR99} Luhman, K.~L.~\& Rieke, G.~H.\ 1999,
\apj, 525, 440 

\bibitem[McLean et al.(1998)]{McLeanetal98} McLean, I. S. et al. 1998, 
\procspie, 3354, 566

\bibitem[Meyer, Calvet, \& Hillenbrand(1997)]{MCH97} Meyer, 
M.~R., Calvet, N., \& Hillenbrand, L.~A.\ 1997, \aj, 114, 288

\bibitem[Montmerle et al.(2000)]{MGTK00} Montmerle, T., Grosso, N., Tsuboi, 
Y., \& Koyama, K. 2000, \apj, 532, 261

\bibitem[Motte, Andre, \& Neri(1998)]{MAN98} Motte, F., 
Andre, P., \& Neri, R.\ 1998, \aap, 336, 150

\bibitem[Ostriker \& Shu(1995)]{OS95} Ostriker, E.~C.~\& 
Shu, F.~H.\ 1995, \apj, 447, 813 

\bibitem[Patterson(1994)]{P94} Patterson, J.\ 1994, \pasp, 106, 209

\bibitem[Shu et al.(1994)]{SNOWRL94} Shu, F., Najita, J., 
Ostriker, E., Wilkin, F., Ruden, S., \& Lizano, S.\ 1994, \apj, 429,
781

\bibitem[Siess, Dufour, \& Forestini(2000)]{SDF00} Siess, L., 
Dufour, E., \& Forestini, M.\ 2000, \aap, 358, 593.

\bibitem[Siess, Forestini, \& Bertout(1999)]{SFB99} Siess, 
L., Forestini, M., \& Bertout, C.\ 1999, \aap, 342, 480

\bibitem[Stahler, Shu, \& Taam(1980)]{SST80} Stahler, S.~W., 
Shu, F.~H., \& Taam, R.~E.\ 1980, \apj, 242, 226 

\bibitem[Stassun et al.(1999)]{SMMV99} 
Stassun, K.~G., Mathieu, R.~D., Mazeh, T., \& Vrba, F.~J.\ 1999, \aj, 117, 
2941

\bibitem[Stassun et al.(2001)]{SMVMH01} Stassun, K.~G., 
Mathieu, R.~D., Vrba, F.~J., Mazeh, T., \& Henden, A.\ 2001, \aj, 121, 1003

\bibitem[Tokunaga (2000)]{Tok00} Tokunaga, A. T. 2000, in {\it Allen's
Astrophysical Quantities Fourth Edition}, ed. A. N. Cox (New York: 
Springer), 143 

\bibitem[Tout, Livio, \& Bonnell(1999)]{TLB99} Tout, C.~A., 
Livio, M., \& Bonnell, I.~A.\ 1999, \mnras, 310, 360

\bibitem[Tsuboi et al.(2000)]{TIKGM00} Tsuboi, Y., Imanishi, K., Koyama,
K., Grosso, N., \& Montmerle, T.\ 2000, \apj, 532, 1089

\bibitem[Wallace \& Hinkle(1996)]{WH96} Wallace, L., \& Hinkle, K. 1996, 
\apjs, 107, 312

\bibitem[Wilking et al.(1989)]{WLY89} Wilking, B. A., Lada, C. J., \&
Young, E. T. 1989, \apj, 340, 823
 
\end{thebibliography}
\end{document}